\documentclass[aps,pre,twocolumn,showpacs,groupaddress]{revtex4}
\usepackage{amsmath}
\usepackage{amsfonts}
\usepackage{amssymb}
\usepackage{epsfig}
\usepackage{graphicx}
\newcommand{\gdot}[0]{\dot{\gamma}}
\newcommand{\Pe}{P_{\rm e}}
\renewcommand{\phi}{\varphi}
\newcommand{\odif}[2]{\frac{\partial #1}{\partial #2}}

\newcommand{\be}{\begin{equation}}
\newcommand{\ee}{\end{equation}}
\newcommand{\ba}{\begin{align}}
\newcommand{\ea}{\end{align}}
\newcommand{\ave}[1]{\langle {#1} \rangle}
\usepackage{color}

\begin{document}

\title{Yield stress in amorphous solids: A mode-coupling theory analysis}

\author{Atsushi Ikeda}
\affiliation{Laboratoire Charles Coulomb, UMR 5221, CNRS and Universit\'e
Montpellier 2, Montpellier, France}

\author{Ludovic Berthier}
\affiliation{Laboratoire Charles Coulomb, UMR 5221, CNRS and Universit\'e
Montpellier 2, Montpellier, France}

\date{\today}

\begin{abstract}
The yield stress is a defining feature of amorphous materials which
is difficult to analyze theoretically, because it stems 
from the strongly non-linear response of an arrested solid 
to an applied deformation. Mode-coupling theory predicts the 
flow curves of materials undergoing a glass transition, and thus
offers predictions for the yield stress of amorphous solids. 
We use this approach to analyse several classes of disordered 
solids, using simple models of hard sphere glasses, soft glasses, and 
metallic glasses for which the mode-coupling predictions can be directly 
compared to the outcome of numerical measurements. The theory 
correctly describes the 
emergence of a yield stress of entropic nature in hard sphere glasses, 
and its rapid growth as density approaches random close packing 
at qualitative level. By contrast, the emergence of solid behaviour in 
soft and metallic glasses, which originates from direct 
particle interactions is not well described by the theory. 
We show that similar shortcomings arise in the description
of the vibrational dynamics of the glass phase at rest. We discuss 
the range of applicability of mode-coupling theory to understand 
the yield stress and non-linear rheology of amorphous materials.
\end{abstract}

\pacs{62.20.-x, 83.60.La, 83.80.Iz}


\maketitle


\section{Introduction}

The yield stress is a defining characteristics of amorphous solids
which represents a robust 
mechanical signature of the emergence of solid behaviour
in many atomic, molecular and soft condensed materials undergoing 
a transition between fluid and solid states~\cite{barnes,coussot2}. 
From a physical viewpoint, the existence of a yield stress implies that 
the material does not flow spontaneously unless a driving force 
of finite amplitude is applied, which represents a very intuitive 
definition of `solidity'. While properly defining and measuring a 
yield stress remains a debated experimental issue~\cite{bonn}, we will 
study simple model systems where the yield stress can be 
unambiguously identified 
as the shear stress $\sigma$ measured in steady 
state shear flow, in the limit where 
the deformation rate $\gdot$ goes to zero, 
\be 
\sigma_Y = \lim_{\gdot \to 0} \sigma(\gdot).
\ee
As such, the yield stress measures a strongly nonlinear 
transition point between flowing states for $\sigma >
\sigma_Y$ and arrested states when $\sigma < \sigma_Y$. 
In this work, we wish to analyze the dependence of $\sigma_Y$ 
upon external control parameters, such as temperature $T$, 
packing fraction, $\phi$, in a wide range of disordered materials. 
Therefore, our work differs from most rheological studies 
of glassy materials which usually describe a set 
of flow curves, $\sigma = \sigma(\gdot)$, for a specific material. 

Dense amorphous particle packings represent a broad class of 
solids possessing a yield stress, which typically emerges
when either temperature is lowered across the glass transition 
temperature $T_g$ in atomic and molecular glasses (such as metallic 
glasses), or when the packing fraction is increased in colloidal hard spheres
and soft glassy materials (such as emulsions 
and soft colloidal suspensions)~\cite{binder,rmp}. 
Of course, the range of materials 
displaying a measurable yield stress is much broader~\cite{barnes}, but we
restrict ourselves to dense particle systems with a disordered,
homogeneous structure, leaving aside systems like colloidal gels
or crystalline and polycrystalline structures.  
 
While our emphasis is mostly on atomic and colloidal systems,
we also include in our discussion materials such as 
foams and noncolloidal soft suspensions, where solidity emerges 
upon compression at the jamming transition, but for which thermal fluctuations 
play a negligible role~\cite{reviewmodes}. 
While the yield stress in jammed solids 
results from the emergence of a mechanically stable 
contact network between particles rather than a glass 
transition~\cite{vhjpcm},  
it was recently demonstrated that the interplay between
glass and jamming transitions can be experimentally relevant 
for the rheology of soft colloidal systems as well~\cite{ikeda}. 
In particular, 
we have shown that the yield
stress of soft repulsive particles displays a very rich behaviour 
as both $T$ and $\phi$ are varied~\cite{ikeda}, and we suggested 
that this is relevant to describe materials such as concentrated 
emulsions~\cite{softmatter} (see also Ref.~\cite{mason}). 

From the modelling point of view, the complex rheology of amorphous 
yield stress materials is often described using simplified or coarse-grained
descriptions that assume from the start
the existence of a yield stress, and study the 
response of the solid to the imposed 
flow~\cite{picard,vdb,jagla,kirsten1,bocquet}. 
Fewer theoretical approaches can describe both
the emergence of a yield stress together with the rheological 
consequences~\cite{sgr,hb,BBK}, as they must then also in principle provide 
a faithful description of the glass or jamming transitions, 
which represent theoretical challenges on their own~\cite{rmp}.
Therefore it should be clear that predicting the 
temperature and density evolution of the yield stress 
across a broad range of materials is much more demanding
than  studying the qualitative evolution of a set of flow
curves. Thus, we hope our study 
will motivate further theoretical developments
to reach this goal.

The mode-coupling theory (MCT) of the glass transition
was first developed in the context of 
the statistical mechanics of the liquid state to account for the 
dynamical slowing down observed in simple fluids approaching the glass 
transition~\cite{gotze}, 
but it has also deep connections to the random first order transition 
theory of the same problem, that are well understood~\cite{rmp}. 
While initially 
thought as a theory for the glass transition, it is now recognized 
that MCT can make relevant 
predictions for time correlation functions
for the initial 2-3 decades of viscous slowing down. Interestingly, 
this time window is very relevant for experiments 
performed in colloidal systems and in computer simulation studies. This
explains why the theory continues to be developed as of today, 
and in particular why its extensions to account for the driven 
dynamics of glasses have experimental 
relevance~\cite{kuni1,kuni2,fuchs1,fuchs2,fuchs3}.
Many specific aspects of the theory have received 
numerical and experimental attention in recent years~\cite{kuni2,fuchs4}, 
but a systematic exploration of the yield stress behaviour
has, to our knowledge, not been performed.

To explore different types of materials while 
keeping the possibility of a direct comparison to theorical 
predictions, we concentrate on simple model systems which can 
be both efficiently studied in computer simulations 
to obtain direct measurements of the yield stress, and
can also be studied within a mode-coupling approach.
Because the static structure of the fluid 
is the only input needed for the theory, measuring 
the structure from computer simulations~\cite{nauroth,gillesmct}
allows us to directly analyse the validity of the theoretical 
predictions, and identify precisely 
the strength, weakness and range of applicability of the theory
to analyze the yield stress of amorphous solids. 
 
In agreement with previous findings, we
observe that for all systems, the theory correctly predicts the emergence of
a finite yield stress as the glass transition is crossed, although it is
difficult to assess quantitatively the detailed predictions 
made by the theory near the `critical' point (because 
the singularity is replaced by a crossover in real systems). For hard sphere 
glasses, the theory accounts qualitatively well for 
both the entropic nature of the solidity (i.e.,
$\sigma_Y \propto k_B T$) and the divergence of the
yield stress as the random close packing density is approached~\cite{mewis}.  
By contrast, we find that the theory fares poorly for systems 
where solidity emerges due direct interparticle interactions 
(i.e., $\sigma_Y \propto \epsilon$, where $\epsilon$ characterizes
the scale of pair interactions) such as soft repulsive 
and Lennard-Jones particles at low temperatures, as theory 
incorrectly predicts that $\sigma_Y \sim k_B T$.  
Our results also show that these shortcomings can 
be traced back to the description of the glass dynamics at rest (i.e.
without an imposed shear flow), rather than to an incorrect 
treatment of the mechanical driving. Therefore, we also 
offer a detailed analysis  of the vibrational dynamics in 
all these models, which is currently the focus 
of considerable attention, in particular in 
colloidal materials~\cite{bonnmode,soft,criticality}. 

The paper is organized as follows.
In Sec.~\ref{models} we introduce our models for hard spheres,
soft and metallic glasses, and the mode-coupling approach we follow
to study the glass dynamics at rest and under flow.
In Sec.~\ref{softrest} we study the vibrational glass dynamics 
of hard sphere and soft sphere glasses at rest.
In Sec.~\ref{softflow} we study the glassy rheology 
of hard sphere and soft sphere models.
In Sec.~\ref{lennard} we repeat the analysis of vibrational 
and rheological properties for Lennard-Jones particles.
In Sec.~\ref{discussion} we discuss our results and offer perspectives
for future research.

\section{Models, methods and mode-coupling theory}

\label{models}

In this section we introduce the models used to 
describe the physics of hard spheres, soft and metallic glasses. 
Then, we describe the simulation methods employed to extract 
vibrational dynamics and the yield stress. 
Finally we present the mode-coupling theory to analyse
both the glass dynamics at rest and its extension to treat 
steady state shear flows. 

\subsection{Model glasses}

In this work, we consider two different model systems. 
To address the physics of hard spheres and soft glasses
we study a system of repulsive harmonic spheres, defined 
by the simple following pairwise potential, 
\be
v_{HS}(r_{ij}) =  \frac{\epsilon}{2} 
(1-r_{ij}/a)^2  \Theta(a - r_{ij}) ,
\ee 
where $\Theta(x)$ is the Heaviside function and $a$ is 
the particle diameter. 

It is well established that harmonic spheres display two 
different regimes when the packing fraction, $\phi$, and 
temperature, $T$, are varied~\cite{tom}. Because of the repulsive 
interaction, harmonic spheres at low temperatures have 
very few overlaps and thus effectively behave, in the limit
of $\epsilon/T \to \infty$ as a hard sphere fluid. In this
regime, the physics of harmonic spheres is controlled by entropic
forces. 
However, this regime can only be achieved if the density is
low enough that configurations with no particle overlap can easily be
found. Upon compression, another regime is entered where particles
have significant overlaps with their neighbors, 
and the system then behaves as a soft repulsive glass.
In this regime, the physics is controlled by the 
energy scale $\epsilon$ of the repulsive forces rather than 
by entropic forces. 
At very low temperatures, the transition between these two distinct glasses 
occurs at the jamming transition~\cite{ohern1}. 
In this paper, our primary goal is not to study the jamming transition 
in detail, but rather to use its existence to study both the 
`entropic' physics of hard spheres and the `energetic' physics
of soft glasses within a single model.

Finally, we use Lennard-Jones particles as a simple model 
for an atomic glass-forming liquid, where the pairwise potential is 
\be 
v_{LJ}(r_{ij}) = 4 \epsilon \left( (a/r_{ij})^{12} - (a/r_{ij})^{6} \right). 
\ee
As we mainly deal with the properties of the glass we use a 
monodisperse Lennard-Jones model. To study also the 
viscous liquid properties, we would need to study 
a system with some size polydispersity (such as
a binary mixture) to prevent crystallization.
Such mixtures are indeed taken as simple models for metallic glasses. 
In this case again, the energy scale $\epsilon$ in the Lennard-Jones 
potential plays a crucial role, as we shall demonstrate.  

\subsection{Computer simulations}

To assess the quality of the mode-coupling theory predictions
we have studied the above models using computer simulations, 
both by producing new data for the present work, and 
by collecting previously published data.
The simulation methods are described in our 
previous publications~\cite{ikeda,criticality,LJ}, and 
so we only give a brief account of these methods. 

To study the vibrational dynamics of the various glass structures, 
we performed Newtonian dynamics simulations. 
We studied the vibrational property of a
single amorphous packing configuration 
at the desired density and temperature, using a very large 
system size~\cite{criticality}.  
To generate the glass configurations, we first prepare a fully 
random configurations, and then
perform an instantaneous quench to very low temperature. We then 
let the system relax until aging effects become negligible,
and purely vibrational dynamics is observed. 
To study glasses at different state points, we heat or cool, we compress
or decompress the initially prepared glass configuration,
followed by a new thermalization.
After the glass structures are obtained, we perform 
production runs. Since we mainly focus on the very low temperatures 
(compared to the glass transition temperature), 
the system lies well inside a metastable state, and particles simply 
perform vibrational motions around their equilibrium positions. 

For numerical simulations of the yield stress of
the harmonic sphere system,  
we performed Langevin dynamics simulations with simple 
shear flow~\cite{ikeda}. The equation of motion is 
\begin{eqnarray}
\xi (\odif{\vec{r}_i}{t} - \gdot y_i \vec{e}_x) = - \sum_{j=1}^N  
\odif{v(|\vec{r}_i - \vec{r}_j|)}{\vec{r}_i} + \vec{R}_i.  \label{eom}
\end{eqnarray}
Here $\vec{r}_i$ represents the position of particle $i$, 
$y_i$ its $y$-component, and  $\vec{e}_x$ the unit vector 
along the $x$-axis. 
The damping coefficient, $\xi$, and the random force, $\vec{R}_i(t)$, obey 
the fluctuation-dissipation relation:
$\ave{\vec{R}_{i,\alpha}(s) \vec{R}_{j,\beta}(s')} = 2 k_B T \xi \delta_{ij} 
\delta_{\alpha\beta} \delta(s-s')$. 
We apply Lees-Edwards periodic boundary conditions. 
We performed sufficiently long simulations at the desired temperature, 
density and shear rate, 
and analyzed their steady state stress measured via the standard 
Irving-Kirkwood formula. The yield stress is typically extracted 
from fitting the steady state flow curves at a given
state point using a phenomenological Herschel-Bulkley
law, $\sigma(\dot{\gamma}) = \sigma_Y + a \dot{\gamma}^n$,
where $a$ and $n$ are additional fitted parameters. 

Because the yield stress of the Lennard-Jones model
has been measured in a number of studies for the case of a well-known
binary mixture~\cite{yieldLJ,yieldLJ2}, we gather these literature data 
as a proxy for the yield stress of 
the monocomponent system. Since our discussion of these data
is mainly qualitative, the differences between both systems
have no impact for the present work.

For both harmonic and Lennard-Jones spheres, we 
use $a$ and $\epsilon/k_B$ as the units of the length and temperature. 
For the time unit, $a(m/\epsilon)^{1/2}$ and $a^2 \xi /k_BT$ are used 
in the inertial dynamics (for vibration) and overdamped dynamics 
(for rheology), respectively, where $m$ is the particle mass.  
We will carefully discuss the appropriate stress scales when needed. 

\subsection{Mode-coupling theory (MCT) of the glass transition}

We present the basic mode-coupling equations
allowing us to describe the dynamics of glassy liquids and 
glasses at rest. The mode-coupling theory (MCT)~\cite{gotze} 
of the glass transition 
can be expressed as a closed set of equations for 
the intermediate scattering functions $F(\vec{k},t) = N^{-1} 
\ave{\rho(\vec{k},0)^*\rho(\vec{k},t)}$.  
Here, $\rho(\vec{k},t) = \sum_i e^{i\vec{k} \cdot \vec{R}_i(t)}$ is the 
instantaneous density field and 
$\vec{R}_i(t)$ is the $i$-th particle position at time $t$. 
The central equation of the MCT is 
\begin{eqnarray}
 \begin{aligned}
\Omega^{-2}(k) \ddot{F}(k,t) + F(k,t) + \int^t_0\!\! ds \ M(k,t-s) 
\dot{F}(k,s) = 0, 
 \end{aligned}
 \label{mctf}
\end{eqnarray}
where $\Omega(k)= \sqrt{k_B T k^2/mS(k)}$ is the frequency term 
associated with acoustic waves, and   
$S(k) = F(k,t=0)$ is the static structure factor.
The memory kernel $M(k,t)$ is given by  
\begin{eqnarray}
 \begin{aligned}
M(k,t) = \frac{\rho S(k)}{2k^2} \int\!\! \frac{d\vec{q}}{(2 \pi)^3}
V(\vec{k},\vec{q},\vec{k}-\vec{q}) F(q,t) F(|\vec{k}-\vec{q}|,t),
 \end{aligned}
  \label{memf}
\end{eqnarray}
with the vertex term 
\begin{eqnarray}
 \begin{aligned}
V(\vec{k},\vec{q},\vec{p}) = \{ \vec{k}\cdot\vec{q}c(q) + \vec{k} 
\cdot\vec{p}c(p)\}^2/k^2. 
 \end{aligned}
  \label{memf2}
\end{eqnarray}
Here, $c(k)=\{1-1/S(k)\}/\rho$ is the direct correlation 
function~\cite{hansen}. 

From the intermediate scattering function, we can also obtain 
various incoherent correlation functions in the framework of the MCT.   
Consider a tagged particle located at $\vec{R}(t)$, and the 
associated density field $\rho_s(\vec{k},t) = e^{i \vec{k} \cdot \vec{R}(t)}$. 
The MCT equations for the self intermediate scattering function 
$F_s(k,t) = \ave{\rho_s(\vec{k},0)^*\rho_s(\vec{k},t)}$ 
have the same structure as Eq.~(\ref{mctf}), but with the 
frequency term now given by 
$\Omega_s(k)=\sqrt{k_BT k^2/m}$ instead of $\Omega(k)$, 
and with the self memory kernel 
\begin{equation}
\begin{aligned}
M_s(k,t) 
\!= \!\frac{\rho}{k^2}\! \int\!\!\frac{d\vec{q}}{(2 \pi)^3}
\left\{\! \frac{\vec{k} \cdot \vec{q}}{k}c(q)\! \right\}^2 \!\!
F_s(q,t) F(|\vec{k}-\vec{q}|,t),
\end{aligned}
\label{mems}
\end{equation}
instead of $M(k,t)$. 
The MCT equation for the mean-squared displacement, $\Delta^2 (t) = 
\ave{|\vec{R}(t) - \vec{R}(0)|^2}$, can also be obtained:
\begin{eqnarray}
 \begin{aligned}
\frac{m}{k_BT} \ddot{\Delta}^2(t) - 6 + \int^t_0\!\! ds \ M_d(t-s) \dot{\Delta}^2(s) = 0,
 \end{aligned}
 \label{mctd}
\end{eqnarray}
where  
\begin{equation}
\begin{aligned}
M_d(t) 
\!= \!\frac{\rho}{6 \pi^2}\! \int\!\! dk \ k^4 c(k)^2 F(k,t) F_s(k,t).  
\end{aligned}
\label{memd}
\end{equation}

The set of MCT equations describes the time evolution of 
the correlation functions $F(k,t)$, $F_s(k,t)$, and $\Delta^2 (t)$. 
The MCT equations have been applied to various model systems including 
the two models studied in this work, 
harmonic spheres and Lennard-Jones 
particles~\cite{nauroth,beng,gillesmct,harmmct}. 
In both cases, the theory predicts an ideal glass transition line in the
$(T, \phi)$  phase diagram.
At high temperature and low densities, $F(k,t)$ and $F_s(k,t)$ 
relax to zero and $\Delta^2(t)$ becomes diffusive, $\Delta^2 (t) \propto 
t$, at long time. 
However when the temperature is decreased and the density is increased, 
the system may enter the non-ergodic glass phase, where 
the long time limits of $F(k,t)$ and $F_s(k,t)$ are positive, and 
the limit of $\Delta^2(t)$ is finite. 

To characterize vibrational dynamics in the glass phase, 
we focus on the mean-squared displacement $\Delta^2 (t)$. 
We compare $\Delta^2 (t)$ obtained from MCT with the direct 
numerical measurements. 
Since the MCT equation of the mean squared displacement 
Eqs.~(\ref{mctd}) and ~(\ref{memd}) 
depend on the collective and self intermediate scattering functions, 
we first need to solve the full MCT equations Eq.~(\ref{mctf}) for these 
correlation functions. 
This requires the static structure factor $S(k)$ as a sole input. 
We use the `exact' $S(k)$ directly obtained from 
the simulations at each state point.
For the numerical integration of Eqs.~(\ref{memf}), (\ref{mems}), 
and (\ref{memd}), we employed equally spaced grids $N_k$ with a 
grid spacing $\Delta k$. 
We use large enough $N_k \Delta k$ and small enough $\Delta k$ to 
be independent from the choice of these parameters. 

Integrating the MCT equations for very low $T$ very
close to the jamming transition for harmonic spheres 
required an usually large number of wavevectors $N_k$, 
as the static structure develops singularities. We made 
sure that all our results are well converged and depend 
only weakly on the numerical integration. We discuss this issue
in more detail in Sec.~\ref{staticjamming}. 
 
\subsection{Mode-coupling theory under shear flow}

In the past decade, the mode-coupling theory of the glass transition
has been extended to study systems under shear 
flow~\cite{kuni1,kuni2,fuchs1,fuchs2}. 
In this work, we follow the approach developed in 
Refs.~\cite{fuchs1,fuchs2}. 
The theory describes a system that is subjected to shear flow at $t=0$, 
and predicts how the system reaches a steady state. 
As before, 
it only requires the static structure factor at rest as an input, 
and gives properties of the steady state under shear flow as output, 
from which we can deduce the value of the yield stress. 

The theory again takes the form of a closed set of equations for
the transient intermediate scattering function, 
$F_t (\vec{k},t) \equiv \ave{\rho (\vec{k},0)^* \rho (\vec{k}(t),t)}$. 
This function is the extension of $F(k,t)$ 
to describe the transient dynamics of the system, 
where the shear flow is applied at $t=0$. The so-called 
advected wave vector
$\vec{k}(t)$ is given by $\vec{k}(t) = \vec{k} - 
\dot{\gamma} k_x \vec{e}_y t$, 
which takes into account the affine advection of density 
fluctuations by the shear flow. 
The central equation of the theory is very similar to the 
usual MCT equation, Eq.~(\ref{mctf}), 
except that the transient correlation function becomes the
unknown function. 
In practice, however, the equations become very difficult to solve because 
the correlation functions are anisotropic, due to the external flow, 
and we cannot perform the circular integral before solving the equations. 

To avoid this problem, we employ the approximation called 
`isotropically sheared model'~\cite{fuchs1}, 
where an isotropic approximation is applied to all correlation functions 
and advected wavevectors. 
In this approximation, the central equation is 
\begin{eqnarray}
 \begin{aligned}
\Gamma^{-1} (k) \dot{F}_t(k,t) +  F_t (k,t) + \int^t_0\!\! ds 
\ M_t(k,t-s) \dot{F}_t(k,s) = 0, 
 \end{aligned}
 \label{mctt}
\end{eqnarray}
where $\Gamma(k) = k_BT k^2/\xi S(k)$ is the damping term and 
$M_t(k,t)$ is the memory kernel given by: 
\begin{eqnarray}
 \begin{aligned}
M_t(k,t) = \frac{\rho S(k)}{2k^2} \int\!\! \frac{d\vec{q}}{(2 \pi)^3}
V_t(\vec{k},\vec{q},\vec{k}-\vec{q},t) F_t(q,t) F_t(|\vec{k}-\vec{q}|,t),
 \end{aligned}
  \label{memt}
\end{eqnarray}
with the vertex term 
\begin{eqnarray}
 \begin{aligned}
V_t(\vec{k},\vec{q},\vec{p},t) = \{ \vec{k}\cdot\vec{q}c(q) 
+ \vec{k} \cdot\vec{p}c(p)\} \\ 
\{ \vec{k}\cdot\vec{q}c(q(t)) + \vec{k} \cdot\vec{p}c(p(t))\} /k^2. 
 \end{aligned}
  \label{vtxt}
\end{eqnarray}
Here $k(t) = k (1 + (\dot{\gamma} t)^2/3)^{1/2}$ is the length of 
the advected wave vector. 

The MCT equations in Eq. (\ref{mctt}) become closed when 
the density, temperature and shear rate are specified and 
the structure factor $S(k)$ for the system at rest are given. 
Once the equation is solved, the time evolution of the transient 
intermediate scattering function $F_t(k,t)$ is obtained. 
Using this correlation function, the shear stress at the desired state point 
can be calculated through 
\begin{eqnarray}
 \begin{aligned}
\sigma = \frac{\dot{\gamma} k_BT \rho^2}{60 \pi^2} \int^{\infty}_0\!\! dt 
\int^{\infty}_0\!\! dk \ 
\frac{k^5 c'(k) c'(k(t))}{k(t)} F_t(k(t),t)^2,
 \end{aligned}
 \label{stress}
\end{eqnarray}
where $c'(k)$ is the derivatives of $c(k)$. 
To solve the equation, we use the same technique as before, 
and again take $S(k)$ as obtained from the simulations. 

\section{Dynamics of hard spheres and soft glasses at rest}

\label{softrest}

We study the vibrational dynamics of 
hard sphere and soft glasses using the harmonic sphere model
in two different density regimes.
Numerical simulations of the vibrational dynamics of this model 
in a wide temperature and density range were reported
before~\cite{criticality}, and we simply summarize the main results. 
We then present the MCT predictions from Eqs.~(\ref{mctf}, \ref{mctd}). 

\subsection{Mean-squared displacement}

\begin{figure}
\psfig{file=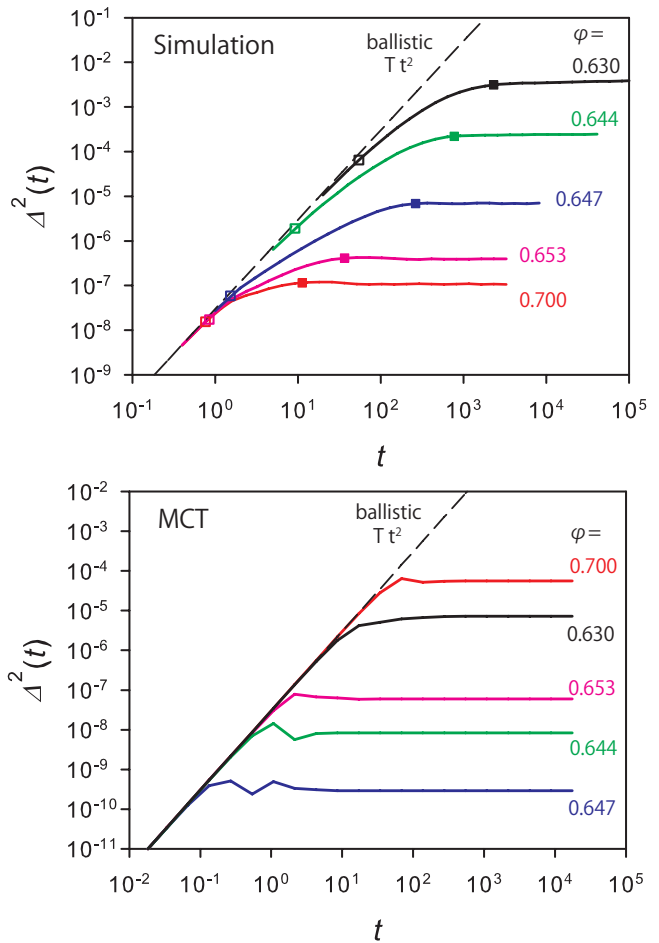,width=8.5cm,clip}
\caption{\label{fig1} 
Top: Time dependence of the mean-squared displacements (MSD) obtained 
from simulation of harmonic spheres at constant temperature, $T=10^{-8}$,
for volume fractions ranging from above to below the 
jamming density $\phi_J \approx 0.647$. 
Open squares indicate the microscopic time scale $\tau_0$ where 
dynamics deviates from ballistic behavior. Filled squares indicate 
the time scale $t^\star$, which marks convergence of the MSD to its 
long-time plateau value. Both time scales decrease with $\phi$, but
their ratio is maximum near $\phi_J$. Bottom: Time dependence of the 
MSD predicted by the MCT at the same state points. There 
is no decoupling between $\tau_0$ and $t^\star$ near $\phi_J$, 
and the plateau value has a non-monotonic density dependence.} 
\end{figure}

We first review the simulation results for the mean-squared 
displacements (MSD). 
The top panel of Fig.~\ref{fig1} shows the time evolution of 
the MSD $\Delta^2 (t)$ at the temperature $T=10^{-8}$ and several densities 
across the jamming density. 
For all densities, $\Delta^2 (t)$ shows ballistic behavior 
$3T t^2$ at very short time, while it approaches a plateau
in the long-time limit. As density increases, 
this plateau value decreases, which shows that 
compressing particles reduces drastically the 
spatial extent of their thermal vibrations, which is physically
expected. 

A closer look at the time dependence of the MSD 
reveals a very interesting behavior in the vicinity of the jamming transition. 
To this end, it is useful to introduce 
the microscopic time scale, $\tau_0$, which coincides with the 
moment where the MSD starts to deviate from its short-time 
ballistic behavior. This timescale $\tau_0$ is  
indicated by open squares in Fig.~\ref{fig1}. Physically, 
it means that particles do not feel their environment for 
$t < \tau_0$. A second relevant timescale, $t^\star$, characterizes
the time dependence of the MSD. It corresponds roughly 
to the timescale at which the MSD reaches its plateau value. 
This corresponds to the time it takes to the particles 
to fully explore their `cage'. This second timescale 
is  indicated by the filled squares in Fig.~\ref{fig1}. 
The precise definitions of these timescales can be found in our 
previous work~\cite{criticality}.

Clearly, while both $\tau_0$ and $t^\star$ decrease when the system is 
compressed, their ratio  evolves in a striking non-monotonic manner
with density, with a maximum occurring very close to $\phi \approx \phi_J$. 
This observation means that, when measured in units of the microscopic
time scale $\tau_0$, vibrations occur over a timescale $t^\star$ 
that is very large near $\phi_J$, but decreases as the packing fraction 
moves away from $\phi_J$ on both sides of the jamming 
transition. This is closely related to the emergence of
dynamic criticality~\cite{criticality} or 
soft modes~\cite{reviewmodes} as the jamming transition 
is approched, $|\phi - \phi_J| \to 0$ and $T \to 0$, with 
clear signatures in the vibrational dynamics at finite temperatures.  

We now compare these results to the MCT predictions
deduced after feeding the MCT equations with the `exact' 
static structure factor $S(k)$ measured in the computer simulations 
at the state points represented in Fig.~\ref{fig1}. 
First, we find that the solution of the MCT equations
corresponds to glassy states, for which
the long time limit of all correlation functions is finite. 
The bottom panel of Fig.~\ref{fig1} shows the MCT results
for the MSD $\Delta^2 (t)$ at the same state points 
as in the top panel. These results show similarities and 
differences with the simulation results. 

The basic time dependence of $\Delta^2 (t)$ is similar to the 
simulation results. The MSD 
show an initial ballistic regime at very short time, and they all
reach a plateau at long time. A first difference 
with the simulations is that the density dependence of this plateau height 
decreases with compression in the hard sphere regime, but 
increases with density above the jamming density, 
which is at odds with the numerical results. 
Regarding the details of the time dependence, the MCT solution 
predicts that the time scales $\tau_0$ and $t^\star$ 
evolve together with a ratio $t^\star / \tau_0$ that
is roughly independent of density. 
There is therefore no separation between microscopic and 
long time scales in these results, and the dynamic criticality 
of the jamming transition is not reproduced by the 
theory. 

This failure is perhaps not too surprising 
as the initial theory was not devised to treat the jamming 
problem. However, we notice that soft modes are directly related to 
clear signatures in the pair correlation function $g(r)$
at short separation $r \approx a$, which we introduced in the 
dynamic equations to produce the results in Fig.~\ref{fig1}.    
These results indicate, however, that this is not enough 
to reproduce the dynamics observed numerically. 

\subsection{Evolution of the Debye-Waller factor}

From the time depencence of the MSD, 
we can extract the long-time limit, 
\be
\Delta^2 (\infty) = \lim_{t \to \infty} \Delta^2 (t),
\ee
which is called the Debye-Waller (DW) factor. We perform 
a quantitative analysis of its evolution over a  wide range of 
temperatures and densities. 

\begin{figure}
\psfig{file=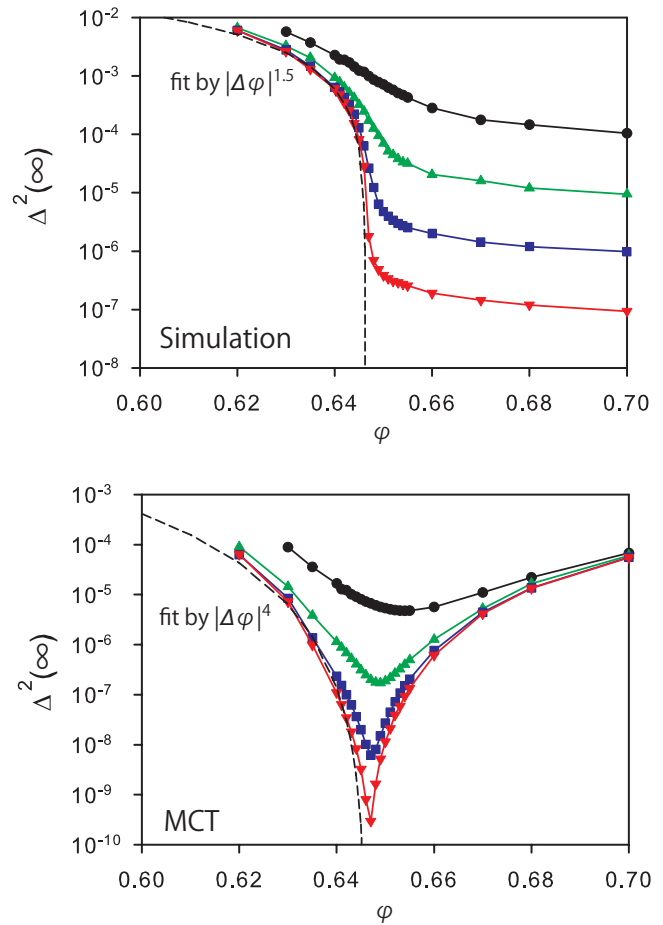,width=8.5cm,clip}
\caption{\label{fig2} 
Volume fraction dependence of the long-time limit of the MSD from 
simulation (top) and from the MCT solution (bottom). 
Different curves correspond to different temperatures from $T=10^{-5}$ 
to $10^{-8}$ (from top to bottom). In simulations, the DW factor decreases 
with $\phi$, with a singular drop near $\phi_J$.
By contrats, the predicted DW factor is non-monotonic with a sharp cusp
near $\phi_J$.  The dashed lines indicate power laws for 
the hard sphere regime,  $\phi < \phi_J$.}
\end{figure}

We show in Fig.~\ref{fig2} the density dependence of the DW factors at various 
temperatures, from $T=10^{-8}$ up to $T=10^{-5}$. In this density
regime, the computer glass transition occurs near $T \approx 5 \cdot 10^{-4}$.  
A first qualitative observation is the confirmation 
that for all temperatures, the DW factor decreases upon compression,
indicating that particles have less space to perform vibrations
at large density.

Second, this figure makes very clear the distinction
between the two types of solids obtained on both sides of the jamming density.
For  $\phi < \phi_J$, which we called `hard sphere glass', 
the DW factor becomes independent of the temperature at low $T$
and is uniquely controlled by $\phi$. In this regime, 
particles are separated by a finite gap 
at very low temperatures, and they can explore this free volume 
regardless of the temperature value. 
On the other hand, when $\phi > \phi_J$, the DW factor is 
proportional to the temperature at low $T$.  
This corresponds to the situation where particles are vibrating 
in an energy minimum created by their neighbors. This
temperature dependence simply corresponds to the low-temperature 
harmonic limit where equipartition of the energy 
yields $\Delta^2(\infty) \propto k_B T$. This is the regime 
we called `soft glass'.
 
The final observation is that upon lowering the temperature,
the density dependence of the DW factor becomes singular on both sides
of the transition, reflecting the emergence of the jamming 
singularity in the $T \to 0$ limit.
Approaching the jamming transition from the hard sphere 
side, the DW factor shows a sharp drop, which is well-described 
by $\Delta^2(\infty) \sim (\phi_J - \phi)^{1.5}$. 
On the other hand, approaching jamming from the soft glass side, 
the DW factor diverges as $\Delta^2(\infty) \sim (\phi - \phi_J)^{-0.5}$. 

These two critical divergences are in fact directly related 
to the slowing down of the vibration discussed 
above~\cite{criticality,brito,leche}. 
To see this, it is useful to define a microscopic length scale $\ell_0$ 
associated to the microscopic time scale $\tau_0$ discussed above,
through $\ell_0 = \sqrt{T} \tau_0$. 
Notably, this length scale is vanishing as jamming is approached 
from the hard sphere side, $\ell_0 \propto (\phi_J - \phi)$, simply
reflecting the vanishing of the interparticle gap. On the soft
sphere side, $\ell_0$ is not singular. Note that the amplitude of
the vibrations quantified by the DW factor vanishes 
less rapidly than $\ell_0^2$ as $\phi \to \phi_J$, 
reflecting the emergence of `soft modes', i.e. collective vibrational
motion that allow large amplitude vibrations, $\Delta^2(\infty) \gg \ell_0^2$.
By renormalizing the DW factor by the microscopic length scale, 
we obtain the density dependence of the adimensional amplitude of 
the vibrational motion, with
\be
\frac{\Delta^2 (\infty)}{\ell_0^2} \propto |\phi_J - \phi|^{-0.5},
\ee
for both hard sphere and soft glass regimes. 
This analysis shows that the amplitude of (adimensional) 
vibrations diverges as $T \to 0$ and $|\phi - \phi_J| \to 0$.

In Fig.~\ref{fig2} we present the MCT predictions for the DW factor
for the same state points as in simulation.
In the hard sphere regime, $\Delta^2(\infty)$ becomes independent
of temperature as $T \to 0$, in agreement with the simulations.
However, the DW factor also becomes independent of $T$
in the soft glass regime, in contradiction to the numerical
findings. It means that MCT cannot account for the fact that
dynamics in the soft glass is controlled by the amplitude of 
interparticle interactions rather than by entropic effects. 
This finding has consequences for the rheology 
of soft glasses, as discussed below. 
 
Regarding the density dependence, MCT correctly predicts
that the DW factor vanishes as $\phi \to \phi_J$ on the hard sphere side. 
Therefore, MCT is able to capture some of the singular features of the jamming
transition. Mathematically, this is because the structure factor 
used as an input to the MCT dynamical equations becomes itself singular
in this limit, which is responsible for the vanishing of the DW factor.
We shall explore this limit in more detail below, but  
the numerical solution of the MCT equations in Fig.~\ref{fig2}
shows that the resulting DW factor vanishes as 
$\Delta^2 (\infty) \sim (\phi_J - \phi)^4$, i.e. with a power law 
that goes to zero much faster than the numerical observations.
Intriguingly, the exponent 4 in this expression is even larger 
than the naive estimate $\Delta^2 (\infty) \sim \ell_0^2 \sim 
(\phi_J - \phi)^2$. This implies that MCT predicts 
that particles are localized over a lengthscale which is much smaller
than the interparticle gap, which is not very physical. 
The second unphysical finding is the overall density 
dependence which is roughly symmetric on both sides 
of the jamming density, with the development of a sharp cusp near 
$\phi_J$ as $T \to 0$ resulting from an incorrect treatment 
of the soft glass dynamics. 

\subsection{MCT predictions near the jamming transition}

We now clarify analytically the nature 
of the MCT predictions near jamming, namely 
that $\Delta^2 (\infty)$ vanishes on both sides of $\phi_J$ with the
same exponent. 
We first analyze the characteristic features of the static structure factor, 
and we then discuss the structure of the MCT equations 
near jamming. 

\subsubsection{Static structure factor near jamming}

\label{staticjamming}

Since the sole input of the MCT equation is the static structure 
factor $S(k)$, 
the predicted singularity of the DW factor must come from 
changes in the structure. The pair structure 
of hard sphere packing near jamming has several 
relevant features~\cite{gr1,gr2,gr3}. 
We find that the MCT equations are most sensitive to the 
simplest of these features, which correspond, in real space, to the 
appearance of a diverging peak at $r = a$ in the pair correlation 
function. Physically, this peak corresponds to the fact that 
at $\phi=\phi_J$ and $T=0$, particles have exactly $z=2d$ contacts, 
i.e. neighbors located at the distance $r=a$. Close to jamming,
$|\phi - \phi_J| \ll \phi_J$, this peak has a finite 
height,  
\be
g_{max} \sim \frac{1}{|\phi - \phi_J|},
\ee
and a finite width $|\phi - \phi_J|$, such that the peak 
turns into a delta function in the limit $\phi \to \phi_J$. 

At first glance, the structure factor $S(k)$ near jamming 
appears not very different from normal fluids~\cite{gr2}. 
It consists of a first diffraction peak near $k = 2 \pi /a$, 
followed by subsequent peaks at larger wavevectors. 
However, the diverging contact peak implies that 
the peaks at large $k$ have an amplitude which decreases 
more slowly than in simple liquids. 
We find that near the jamming density the envelope of the peaks 
of $S(k) - 1$ first decreases as $k^{-1}$, followed
by a crossover to a $k^{-2}$ behavior. 
When $\phi$ gets closer to $\phi_J$, the crossover wavevector 
$k^\star$ between these 
two power laws occurs at larger $k$,  and it 
scales as $k^\star \sim g_{max}$. In summary, we find the following behaviour: 
\begin{eqnarray}
\mbox{peak heights of} \ [S(k)-1] &\approx & \frac{1}{k}, 
\ \ \ \ \ 1 \ll k \ll g_{max}, 
\nonumber \\ 
&\approx & \frac{g_{max}}{k^2},  \ \ \ g_{max} \ll k. 
\label{sofk}
\end{eqnarray}

\begin{figure}
\psfig{file=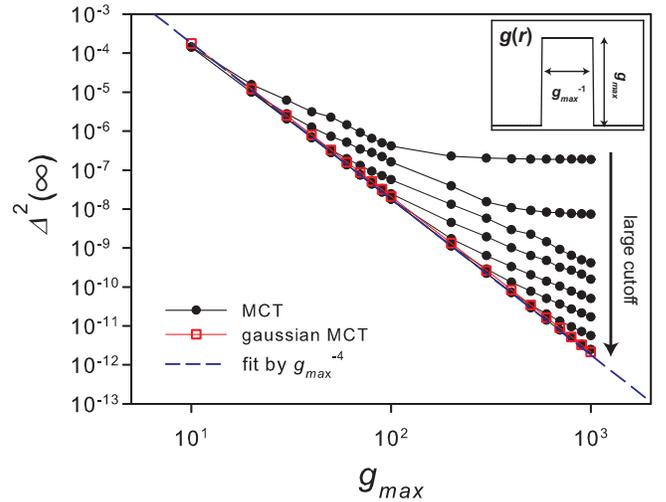,width=8.5cm,clip}
\caption{\label{fig3} 
Evolution of the Debye-Waller factor approaching the jamming transition 
from the hard sphere side, parametrized the maximum 
$g_{max} \sim 1/|\phi_J - \phi|$ of 
the first peak of the pair correlation function.
The DW factor predicted by the full MCT equation Eq.~(\ref{mctd}) 
(filled symbols) converges for a large enough cutoff
to the result obtained from the Gaussian approximated MCT 
equation Eq.~(\ref{mctg}) (open symbols) using  
the simplified pair correlation function shown in the inset.
All solutions agree with $\Delta^2 (\infty) \approx g_{max}^{-4}$, and with the 
full MCT solution in Fig.~\ref{fig2}. }
\end{figure}

To make analytic progress, we introduce a simplified model 
for the pair correlation $g(r)$ near jamming, 
\be
g(r) = g_{max}, \quad {\rm for} \, \, 1 \leq r \leq 1+g_{max}^{-1},
\label{gofr}
\ee 
and $g(r) = 0$ otherwise. 
This model $g(r)$ is illustrated in the inset of Fig.~\ref{fig3}. 
The Fourier transform of this rectangular function can be easily performed,
and provides the scaling behavior of $S(k)$, 
\begin{eqnarray}
S(k)-1 &\approx & \frac{\sin(k)}{k}, \ \ \ \ \ 1 \ll k \ll g_{max}, 
\nonumber \\ 
&\approx & \frac{g_{max}\cos(k)}{k^2},  \ \ \ \ \ g_{max} \ll k, 
\label{sofkm}
\end{eqnarray}
which is essentially the same as Eq.~(\ref{sofk}).
In the limit of the jamming density, $S(k)$ becomes 
$S(k) - 1 \sim \sin (k)/k$, which is exactly the Fourier transform of 
the delta function in three dimensions. 
This means that the model Eq.~(\ref{gofr}) captures the large $k$ behavior 
of the real $S(k)$ correctly. 

We have solved the MCT equations with the Fourier transform of 
Eq.~(\ref{gofr}) as an input for the structure factor. 
In Fig.~\ref{fig3} we show the evolution of the DW factor 
parameterized by the value of $g_{max}$, which diverges as
$\phi \to \phi_J$. We present the results of the numerical 
solution obtained for different values for the wavevector 
cutoff, $N_k \Delta k$, showing that when the numerical solution 
has converged, a perfect agreement is obtained for the evolution of the 
DW factor from the numerically determined structure factor, 
and from the simplified model Eq.~(\ref{gofr}). 
Both MCT solutions, when properly converged, 
result in the scaling behavior 
$\Delta^2 (\infty) \approx g_{max}^{-4} \sim (\phi_J - \phi)^{4}$.
This agreement shows that the MCT solution is dominated by the large 
$k$ behavior of the $S(k)$, Eq.~(\ref{sofk}), 
and therefore is well captured by our simplified model in
Eq.~(\ref{gofr}).

\subsubsection{Analysis of the MCT equation: Gaussian approximation}

To finally analyze the origin of the power law $\Delta^2 (\infty) \approx 
g_{max}^{-4}$, we introduce a simplified version of the MCT 
equation, called Gaussian approximated MCT. 
Assuming that $F(k,t)$ and $F_s(k,t)$ have a Gaussian wavevector 
dependence and that $F(k,t) \approx S(k) F_s(k,t)$,
which is the so-called Vineyard approximation 
(both conditions accurately hold in the full MCT solution), 
the MCT equation can be analytically simplified~\cite{ikedamct}. 
The long-time limit of this simplified MCT equations becomes 
\begin{eqnarray}
 \begin{aligned}
\frac{1}{\Delta^2(\infty )} = \frac{\rho}{6 \pi^2} \int dk \ k^4 c(k)^2 
e^{-2 \Delta^2 (\infty) k^2}.  
 \end{aligned}
 \label{mctg}
\end{eqnarray}
This equation takes $S(k)$ as a sole input (the direct correlation 
$c(k)$ follows directly from $S(k)$) as in the case of the full 
MCT equations. 
We again solve this equation numerically, and show the results 
in Fig.~\ref{fig3} as open squares. 
The solution perfectly agrees with the solution of the full MCT 
equation with the full $S(k)$. 

The advantage of the formulation in Eq.~(\ref{mctg}) is that the asymptotic 
behaviour of the DW factor can now be understood analytically.
Using the behaviour of $S(k)$ in Eq.~(\ref{sofkm}), 
the integrant in Eq.~(\ref{mctg}) becomes 
$k^2 e^{-2 \Delta^2 (\infty) k^2}$ for $k \ll g_{max}$, 
and $g_{max}^2 e^{-2 \Delta^2 (\infty) k^2}$ when $k \gg g_{max}$. 
Here we omitted the square of trigonometric functions 
since they only give constant contributions. 
When $\Delta^2 (\infty) \ll g_{max}^{-2}$, the integral is dominated by the 
contribution from $k \gg g_{max}$. 
This integral can be performed as a Gaussian integral, and this gives 
$\Delta^2 (\infty) \approx g_{max}^{-4}$, which  
also agrees with the assumption $\Delta^2 (\infty) \ll g_{max}^{-2}$, 
and with the observation from the full MCT equation. 

In summary, by simplifying the full MCT treatment with 
the exact $S(k)$ using both a simplified model for $g(r)$ and a Gaussian
approximation of the MCT equations, we can establish analytically
the MCT result $\Delta^2 (\infty) \sim (\phi_J - \phi)^{4}$, which 
is mainly controlled by the large $k$ behaviour of $S(k)$, produced
by the divergence of the contact peak in $g(r)$. 

\subsection{Discussion of the MCT near jamming}

We have unveiled two distinct features of the MCT predictions 
for the DW of harmonic spheres near the jamming transition.

First, we discussed the behaviour in the hard sphere regime,
where a power law vanishing of the DW factor with a large 
exponent is found. We revealed that this power law 
is dominated by the behavior of the static structure factor at large 
wavevector $k^\star \gg g_{max}$. Since $1/k^\star$ represents 
the typical gap between particles, this finding implies 
that the MCT equations are actually controlled 
by lengthscales which are smaller than the typical gap.
This is in clear disagreement with the numerical finding that the 
DW factor corresponds to an amplitude for the vibrations 
that is actually much larger the interparticle gap. 

The second problem is more general, and thus more severe. 
In the soft glass regime,  $\phi > \phi_J$, the predicted 
DW factor not only has the incorrect asymptotic behaviour, 
but it also has incorrect temperature and density dependences. 
This results from the fact that the solution of the MCT is controlled by 
$g_{max}$, while in the soft glass regime the system simply vibrates 
harmonically near the energy minimum. This physics is not captured
by the MCT equations which instead again describe this harmonic
solid as an `entropic' system. This results in the prediction of 
a DW factor that remains finite as $T \to 0$ at large density,
instead of the linear temperature dependence expected in this 
limit. We note that this problem is not specific to harmonic 
spheres and is actually very general for systems with continuous 
pair potentials, as will be shown in Sec.~\ref{lennard} where Lennard-Jones 
particles are considered. 

\section{Hard spheres and soft glasses under flow}

\label{softflow}

In this section, we study the shear rheology of hard sphere 
and soft glasses extending the results in Sec.~\ref{softrest} 
to include shear flow. 
We start with the analysis on the flow curves and then provide 
a more detailed discussion of the yield stress behavior.

\subsection{Flow curves}

\begin{figure}
\psfig{file=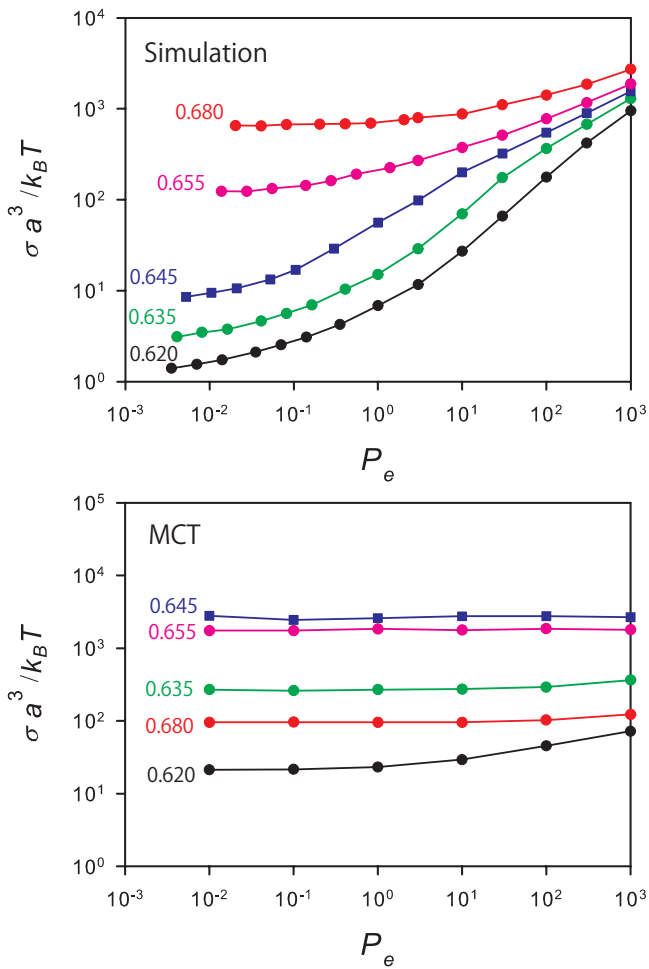,width=8.5cm,clip}
\caption{\label{fig4} 
Top: Flow curves obtained from simulation 
of harmonic spheres at $T=10^{-6}$ and various volume fractions. 
A finite yield stress exists for all $\phi$, which increases
monotonically with the density. The athermal rheology 
of soft repulsive particles near jamming appears at large 
Peclet number, $\Pe > 1$. 
Bottom: The MCT flow curves for the same state points 
as the top panel produce a finite yield stress at all $\phi$
which is maximum near $\phi_J \approx 0.647$, but decreases
with $\phi$ above jamming.}
\end{figure}

We start with a brief review of the simulation results
for the flow curves of harmonic spheres~\cite{ikeda}. 
In the top panel of Fig.~\ref{fig4}, we present several flow curves, 
$\sigma = \sigma(\dot{\gamma})$, 
at low temperature $k_BT /\epsilon = 10^{-6}$ and various densities crossing 
the jamming density $\phi_J$. We use adimensional units
for both the stress scale (using $k_B T/a^3$ as thermal stress unit), 
and for the shear rate (using the Peclet number 
$\Pe = \dot{\gamma} a^2 \xi/ (k_BT)$). 

First, we focus on the slow shear rate regime, $\Pe < 1$. 
All the flow curves show that the stress approaches a constant value, 
the yield stress $\sigma_Y = \lim_{\dot{\gamma} \to 0} \sigma(\dot{\gamma})$. 
The yield stress increases rapidly with increasing the density. 
At lower density in hard sphere regime $\phi < \phi_J$, 
the stress is $\sigma_Y a^3 /k_BT = \mathcal{O}(1)$, indicating the 
entropic nature of the stress, 
and it increases rapidly when the jamming transition 
is crossed, suggesting that it is not controlled by entropic 
forces alone in this regime.  

Next, we focus on the fast shear rate regime, $\Pe > 1$. 
In this regime, the flow curve shows complex and 
interesting behavior~\cite{ikeda}. 
At low density, the flow curve exhibits a crossover between 
strong shear-thinning when $\Pe < 1$ to Newtonian behavior
when $\Pe > 1$. This shows that a system that looks 
solid at low $\Pe$ in fact appears as a fluid when $\Pe$
becomes large, characterized by an `athermal' Newtonian viscosity.
This viscosity increases rapidly with the density, and this
Newtonian regime disappears above the jamming density, 
where it is replaced by the emergence of a finite   
yield stress. 

We solved the MCT equation, Eq.~(\ref{mctt}), at the desired shear rate and 
with the static structure factor obtained from simulation at the desired 
density and temperature, following the same procedure as for the 
mean-squared displacement in the previous section.
The bottom panel of Fig.~\ref{fig4} shows the flow curves
obtained within MCT. As for the numerical results, 
the MCT flow curves at these state points are all approaching 
a finite yield stress at low shear rate, implying that 
MCT correctly predicts that these glass states offer
a finite resistance to shear flow.

In the hard sphere regime, the yield stress increases 
rapidly with density, 
which qualitatively agrees with the numerical observations.
However, the yield stress is found to decrease with density 
in the soft glass regime, in disagreement with the simulation results. 
This incorrect behavior is very similar to the one reported 
for the vibrational dynamics in the previous section, and we will 
argue below that it has the same origin. 

Finally, we focus on the MCT predictions for $\Pe > 1$. 
The MCT flow curves in this regime do not exhibit the interesting 
behavior observed in the numerical simulations. This 
is not very surprising as the MCT under shear flow 
is specifically designed to treat systems controlled by thermal 
fluctuations, which become inefficient when $P_e > 1$. 
This result nevertheless clearly reveals that a
naive extension of the MCT will not be sufficient 
to treat the interesting zero-temperature shear rheology of 
soft particle systems, which is currently the focus of a large
interest~\cite{ikeda,softmatter,olsson,boyer,lerner}.
 
\subsection{Temperature and density evolution of the yield stress}

\begin{figure}
\psfig{file=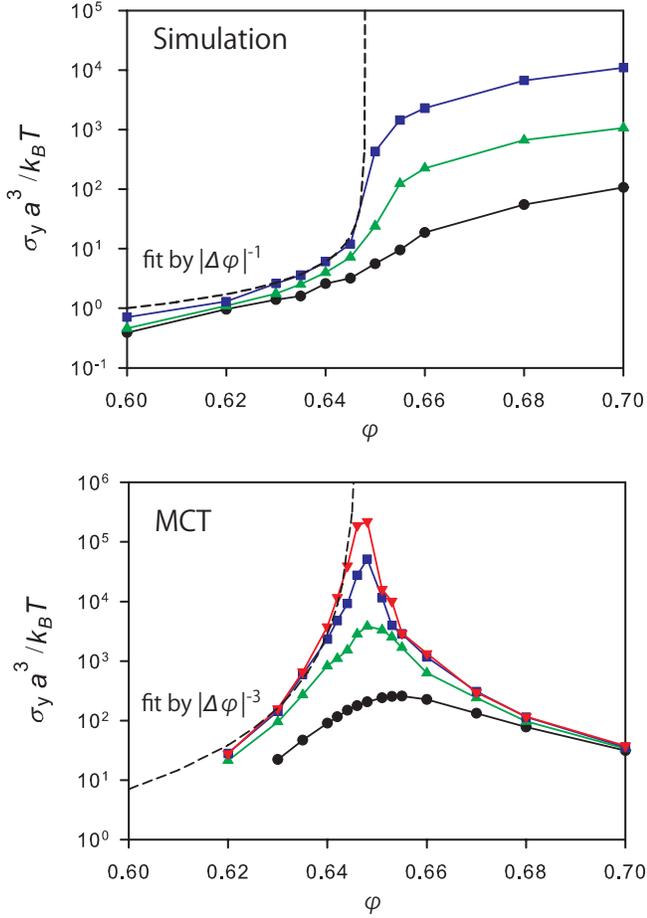,width=8.5cm,clip}
\caption{\label{fig5} 
Top: Volume fraction dependence of the yield stress from the simulation
from $T=10^{-5}$ to $10^{-7}$ (bottom to top). 
The yield stress increases monotonically with the emergence of 
sharp singularities near $\phi_J$ as $T \to 0$. 
Bottom: MCT predictions for the same state points with 
$T=10^{-8}$ added.
The yield stress is non-monotonic with a sharp cusp near $\phi_J$. 
The dashed lines indicate power laws for the hard sphere regimes, 
$\phi < \phi_J$.}
\end{figure}

We finally arrive at the analysis of the yield stress 
in hard sphere and soft glasses. 

We show in Fig.~\ref{fig5} the density dependence of the 
yield stress $\sigma_Y$ measured in the numerical simulation 
of harmonic spheres at various temperatures. These data 
confirm that the yield stress increases monotonically upon compression,
as was observed in the flow curves.  
As in the case of the Debye-Waller factor, 
the temperature dependence of the yield stress is different on both sides 
of the jamming. 
For $\phi < \phi_J$, the entropic nature of the yield stress is obvious 
since it becomes proportional to temperature. 
In the adimensional represention of Fig.~\ref{fig5}, this means 
that $\sigma a^3 /(k_BT)$ becomes uniquely controlled 
by $\phi$ in the hard sphere regime. 

On the other hand, when $\phi > \phi_J$, the nature of the 
yield stress changes from being entropic to being controlled 
by the energy scale governing the particle repulsion, 
i.e. $\sigma_Y \sim \epsilon / a^3$. In the adimensional
representation of Fig.~\ref{fig5}, the data become proportional
to $\epsilon / (k_B T)$.
In this regime, the stress does not originate from thermal collisions
between hard particles, 
but stems from direct interactions between particles interacting with a
soft potential characterized by the energy scale $\epsilon$. 

Having clarified the temperature dependence 
in the two regimes, we turn to the density dependence
which becomes singular around the jamming density when 
temperature becomes small, mirrorring again the behavior
of the DW factor. 
In the hard sphere regime, the yield stress increases rapidly as $\phi_J$
approaches, with 
\be
\sigma_Y \sim \frac{k_B T}{a^3} \frac{1}{(\phi_J - \phi)}.
\ee 
In the soft glass regime at low $T$, 
the yield stress vanishes when the jamming transition 
is approached, with 
\be
\sigma_Y \sim \frac{\epsilon}{a^3} (\phi - \phi_J).
\ee
These two asymptotic behaviors are clearly observed in 
Fig.~\ref{fig5}.
 
The bottom panel of Fig.~\ref{fig5} presents the MCT predictions
for the yield stress for the same state points. The theory 
predicts that the yield stress results from entropic forces on both 
sides of the transition, failing to recognize the change
to the soft glass regime dominated by interparticle forces.
As a result, the theory incorrectly predicts the emergence 
of a cusp as $T \to 0$, with a symmetric divergence of the yield
stress on both sides of $\phi_J$, which is only observed 
on the hard sphere side in the simulations.
 At the quantitative level, the MCT predicts a power law 
divergence on the hard sphere side, 
$\sigma_Y \sim \frac{k_B T}{a^3} (\phi_J - \phi)^{-3}$,
but the exponent 3 differs from the numerical result
although the (entropic) prefactor has the right scaling.

Overall, the degree of consistency between theory and simulation 
for the yield stress is very similar to the 
one deduced from the analysis of the DW factor in the previous
section. In the following section, we rationalize this similarity.

\subsection{MCT predictions near the jamming transition}

We now provide an explanation for the MCT prediction of the 
yield stress divergence $\sigma_Y \propto 
(\phi_J - \phi)^{-3}$ in the hard sphere regime, and of the 
qualitatively incorrect scaling found in the soft glass regime.
To do so, we analyze the 
structure of the MCT equations under shear flow. 

In the MCT framework, the stress is expressed as 
an integral over time and wavevector, see Eq.~(\ref{stress}). 
For the present analysis, it is useful to rewrite the integral as 
\begin{eqnarray}
\sigma = \dot{\gamma} \int^{\infty}_0 dt \ G(t),
\end{eqnarray}
where 
\begin{eqnarray}
G(t) = \frac{k_BT \rho^2}{60 \pi^2} \int^{\infty}_0\!\! dk \ 
\frac{k^5 c'(k) c'(k(t))}{k(t)} F_t(k(t),t)^2,
\end{eqnarray}
is the MCT expression of the transient stress auto-correlation function. 
Using this expression, we can analyze the shear stress in terms of the 
relaxation behavior of $G(t)$. 

\begin{figure}
\psfig{file=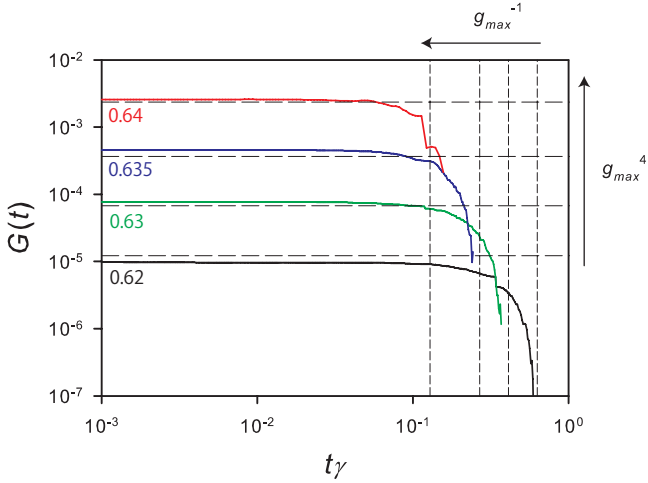,width=8.5cm,clip}
\caption{\label{fig6} 
Time dependence of the stress autocorrelation function $G(t)$ 
obtained from the numerical solution of the MCT equations
under shear flow in the hard sphere regime approaching jamming,
for $T=10^{-6}$ for $\Pe = 10^{-3}$. 
The horizontal and perpendicular dashed lines respectively 
scale as $g_{max}^4$ and $g_{max}^{-1}$, 
in agreement with Eqs.~(\ref{gp}, \ref{ty}).}  
\end{figure}

In Fig.~\ref{fig6}, we show $G(t)$ for $T=10^{-6}$ and several densities 
in the hard sphere regime below jamming. 
The shear rate is fixed at $\Pe = 10^{-3}$, where the stress is 
nearly equal to the yield stress. 
As discussed in the previous section, the yield stress increases rapidly with 
approaching jamming. 
The data in Fig.~\ref{fig6} show that the stress increase 
results from the combination of two different contributions. 
A first factor is the sharp increase of the plateau height of 
$G(t)$ with density. 
The second factor is the decrease of the relaxation time of $G(t)$ with 
increasing the density. We now analyse these two factors separately.

When $t$ is small as $t \dot{\gamma} \ll 1$, the advected wavevector $k(t)$ 
is essentially equal to the wavevector at rest, $k(t_{micro} 
\ll t \ll \dot{\gamma}^{-1}) \approx k$, 
where $t_{micro}$ is the microscopic time to reach the plateau. 
In this case, the sheared MCT equations Eq.~(\ref{mctt}) is 
nothing but the usual MCT equation, Eq.~(\ref{mctf}). 
Thus, the transient intermediate scattering function $F_t(k,t)$ in this regime 
can be accurately approximated by the usual intermediate scattering function 
$F(k,t)$, with no influence from the wavevector advection. 
In this regime, the plateau height of $G(t)$ can be rewritten as 
\begin{eqnarray}
 \begin{aligned}
G_{p} \approx \frac{k_BT \rho^2}{60 \pi^2} \int^{\infty}_0\!\! dk \ 
k^4 c'(k)^2 F(k,t)^2.
 \end{aligned}
\label{modulus}
\end{eqnarray}
This expression is exactly the one provided by MCT 
for the shear modulus of the glass at rest~\cite{nagele}.
Furthermore, the behavior of $c'(k)$ at large $k$ is the same as $c(k)$, 
since $c(k)$ is asymptotically a product of a fast oscillating 
function and a slowly decreasing function of $k$ 
as in Eq.~(\ref{sofkm}), and thus the amplitude of $c(k)$ and $c'(k)$ is 
asymptotically the same. Therefore, Eq.~(\ref{modulus}) 
is also essentially equivalent to the right hand side 
of Eq.~(\ref{mctg}), which enters the expression of the DW factor. 
This means the plateau height behaves as 
\be
G_{p} \sim k_B T g_{max}^{4},
\label{gp}
\ee 
showing that the shear modulus scales with density as 
the inverse of the DW factor, with a temperature prefactor 
revealing its entropic nature.  
In Fig.~\ref{fig6}, we represented dashed lines at levels 
scaling with $g_{max}^4$, which confirm that $G_p$ indeed follows 
Eq.~(\ref{gp}).

The second factor contributing to the scaling of the shear stress
is the relaxation time of $G(t)$. 
In the sheared MCT, the memory function becomes explicitly time dependent 
because of the advection of the wave vectors.
A decoupling between $k$ and the advected $k(t)$ occurs at long time, 
which results in a dephasing of the oscillations of $c(k)$ and 
$c(k(t))$. We have shown in the previous section 
that the MCT integral are dominated by a Gaussian 
contribution $\sim \exp(-2 \Delta^2(\infty) k^2)$, showing 
that we need to consider the decoupling of wave vectors 
for $k \sim 1/\Delta(\infty)$. This occurs after a time 
$t_Y$ such that $k(t_Y) - k = \mathcal{O}(1)$. This 
produces an estimate for the relaxation 
time of the stress autocorrelation function, 
\be
t_Y \dot{\gamma} \sim k^{-1/2} \sim g_{max}^{-1}.
\label{ty}
\ee 
We plot this estimate in Fig.~\ref{fig6} with vertical lines 
scaling with $g_{max}^{-1}$. 
Clearly, these lines agree very well with the relaxation time of $G(t)$
obtained from the numerical resolution of the MCT equations. 
Since we focus on the relaxation dynamics of the system subjected to the 
shear flow starting at time $t=0$, 
$t_Y$ measures the time it takes the glass to yield.
We can therefore identify $\gamma_Y = t_Y \dot{\gamma}$ 
with the {\it yield strain}. 

By combining the MCT prediction for the divergence of the shear 
modulus near jamming as $G_{p} \approx k_BT (\phi_J - \phi)^{-4}$, and 
for the vanishing of the yield strain as $\gamma_Y \approx (\phi_J 
- \phi)^{1}$, we obtain 
the divergence of the yield stress as
$\sigma_Y \approx  G_{p} \gamma_Y \approx k_B T (\phi_J - \phi)^{-3}$.  
This scaling law agrees very well with the numerical solution 
of the MCT equations shown in Fig.~\ref{fig5}, as announced. 
 
\subsection{Discussion of the MCT under flow near jamming}

The above analysis clarifies that the MCT under shear flow 
makes predictions for the yield stress which are direct 
consequences of the behavior obtained from the MCT 
dealing with the glass dynamics at rest.
Within MCT, the yield stress can be expressed as the product of the shear 
modulus and the yield strain, $\sigma_Y = G_{p} \gamma_Y$, 
and the shear modulus in the MCT framework, Eq.~(\ref{modulus}), is 
closely related to the DW factor. 
Thus, the discussions of the MCT predictions near jamming 
for the DW factor and the yield stress are essentially the same. 
In the hard sphere regime, MCT correctly describes 
the entropic nature of the yield stress and its divergence 
as $\phi_J$ is approached, but the critical exponent for the divergence
is too strong. In the soft glass regime, the theory 
incorrectly predicts a scaling of the yield stress with $k_B T$, failing
to detect the direct influence of the particle interactions.

However, we wish to note that  MCT 
provides a new prediction for the yield strain $\gamma_Y$
in the hard sphere regime, $\gamma_Y \approx (\phi_J - \phi)$. 
This is an interesting novel critical behaviour, although the 
predicted value for the associated 
critical exponent is not correct. Indeed, 
in the simulations one has $\sigma_Y \sim 
(\phi_J - \phi)^{-1}$~\cite{ikeda},
while the shear modulus scales with a different exponent, 
$G_p \sim (\phi_J - \phi)^{-1.5}$~\cite{britoepl}. 
This indicates that the yield strain actually vanishes as 
$\gamma_Y \approx (\phi_J - \phi)^{0.5}$.
Note that recent experiments in dense emulsions show 
that the yield strain decreases when the jamming transition is 
approached~\cite{mason}. 

A simple argument can rationalize the 
critical scaling of the yield strain. 
The yielding predicted by the MCT occurs due to the decoupling 
between the wavevector and the advected one at the `relevant' length scale. 
Using the correct value of the interparticle gap for 
this length scale, one directly 
predicts that yielding occurs when the typical gap between neighboring 
particles $\delta \sim |\phi - \phi_J|$ is blurred by the shear 
deformation. (A similar argument was used in Ref.~\cite{criticality}
to discuss thermal effects.) The shear flow causes a transverse displacement 
of particles over a length $\gamma a$, 
and this causes a change in the interparticle distance $\gamma^2 a$. 
Yielding then occurs when $\gamma_Y^2 a \approx \delta$, 
which gives $\gamma_Y \approx (\phi_J - \phi)^{0.5}$, 
as observed numerically.
Note that the argument can be repeated above the jamming 
transition in the soft glass regime.
Also in this regime, there is a mismatch between the scalings of yield stress~\cite{olsson,ikeda}
and shear modulus~\cite{ohern1}, indicating that the yield strain vanishes as 
$\gamma_Y \approx (\phi - \phi_J)^{0.5}$. 
This is again consistent with the idea that the particle overlap 
$\delta \sim |\phi - \phi_J|$ is the relevant length scale. 

\section{Lennard-Jones glass dynamics}

\label{lennard}

In this section, we focus on Lennard-Jones particles, with two main
justifications. First, this allows us to treat a very different
type of material, as Lennard-Jones fluids are often taken as simple
models to study metallic glasses~\cite{binder}. 
A second goal is to investigate 
further the generality of the findings of the previous sections 
concerning the difficulty encountered by MCT in describing 
amorphous materials when solidity emerges from direct, continuous  
interparticle forces. 
We first analyze the structure of the Lennard-Jones glass, 
then its vibrational dynamics, and we finally 
study the yield stress measured under shear flow.

\subsection{Glass structure factor}

To solve the MCT equations under shear flow, we need the glass 
structure factor as an input. We use the 
structure factor measured in low-temperature 
numerical simulations of the monodisperse Lennard-Jones system.
However, since we use a monodisperse system, 
crystallization takes place if we use a temperature which 
becomes too close to the glass transition and diffusive
motion becomes possible. To avoid this problem, 
we need to restrain ourselves on relatively low temperatures. 

To extend our analysis to higher temperatures, 
we implement a second strategy. We use statistical mechanics
to predict the structure factor of the Lennard-Jones fluid
and glass states combining the hyper-netted chain approximation
for the fluid~\cite{hansen}, to the replica approach of Ref.~\cite{mp}
for the glass. While we do not expect this approach to be very 
accurate, it still provides
structure factors that are qualitatively correct down to very low temperatures,
encompassing both fluid and glass states. 
Using this approach, we find that MCT predicts 
a kinetic arrest occurring at $T_{mct} \approx 1.2$, 
while the replica approach yields a Kauzmann transition 
at lower temperature, near $T_K \approx 0.9$. 
Above $T_K$, $S(k)$ is identical to the prediction of the 
hypernetted chain approximation, while below $T_K$ the glass 
structure differs from the liquid state approximation~\cite{mp}. 

The key point of both approaches is that when $T$
becomes smaller than the computer glass transition, 
$S(k)$ rapidly converges towards its $T \to 0$ limit, and 
has actually a very weak temperature dependence in the glass phase.  
By contrast with the jamming point, however, 
$S(k)$ does not develop any kind of singularity even as $T \to 0$.
This directly implies that DW factor and yield stress should 
behave smoothly in the glass phase of the Lennard-Jones system.
 
The reason for this becomes clearer if one focuses on the pair correlation 
function. In the frozen glass state, the distance between any
two particles fluctuates around its average value with a variance 
proportional to $k_B T$. However, since the structure 
is fully amorphous, the {\it spatially averaged} pair correlation
function remains non-singular as $T \to 0$ because the successive correlation 
peaks are broadened by the {\it quenched disorder} imposed by the amorphous
structure.    
 
Therefore, when lowering $T$, there exists a temperature crossover, 
$T_q$, below which the thermal broadening of the peaks in the pair structure 
becomes smaller than the broadening due to the quenched  
disorder. When $T < T_q$, $S(k)$ and $g(r)$ do not depend on 
$T$ anymore, and the solution to the MCT dynamic 
equations remain the same as $T$ is decreased further. We shall see
that MCT yields physically incorrect solutions below $T_q$.  

\subsection{Temperature evolution of the Debye-Waller factor}

\begin{figure}
\psfig{file=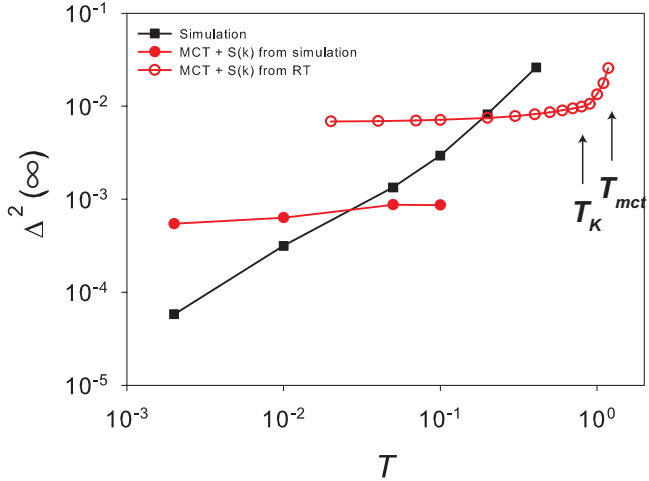,width=8.5cm,clip}
\caption{\label{fig7} 
Temperature dependence of the DW factor obtained from direct
numerical simulation (filled squares) and from MCT
using either the numerically measured structure factor 
(filled circles) or a statistical mechanics approach 
(Replica Theory, open circles). While the simulation results indicate a linear
dependence on $T$, the MCT solution suggests a singular 
$T$-dependence near $T_{mct}$ followed by a rapid saturation to an
unphysical $T$-independent value.}
\end{figure}

We start our analysis of the MCT predictions 
for the Lennard-Jones glass 
with the characterization of the vibrational dynamics. 
We focus on the temperature dependence of the DW factor for a fixed 
number density, $\rho = 1.2$. 

The temperature dependence of the DW factor $\Delta^2 (\infty)$ 
obtained from direct numerical simulations is plotted 
in Fig.~\ref{fig7} together with the results from the MCT solution.
The simulation results (filled square) show that 
the DW factor is proportional to temperature when temperature 
becomes small, which is the same behavior as observed for the 
soft glass in Sec.~\ref{softrest}. This corresponds again to the
limit of the Einstein harmonic solid where the amplitude 
of the vibrations around the average position 
is proportional to $k_B T$, as a direct result of 
equipartition of the energy.
The data in Fig.~\ref{fig7} indicate that this linear behaviour is
obeyed to a good approximation nearly up to the glass transition
temperature.

The MCT analysis performed using the static structure factor 
obtained from simulation is shown with filled circles, 
which indicate that the DW factor 
is nearly independent of temperature in this regime. 
This result follows from the above discussion of the static structure
which is also temperature independent, but clearly contradicts
the numerical simulations. 
This discrepancy is in fact equivalent to 
the findings obtained in the soft glass regime of harmonic spheres. 

Finally, using the analytic structure factor, we can follow
the DW factor to higher temperatures and describe 
the emergence of a finite DW factor, $\Delta^2_c$, 
at the predicted mode-coupling transition, $T_{mct} = 1.2$. 
The theory then predicts 
an abrupt temperature dependence characterizing by a 
square root singularity~\cite{gotze}, $\Delta^2(\infty) \sim \Delta^2_c - a 
\sqrt{T_{mct} -T}$, where $a$ is a numerical prefactor.
However, the temperature evolution of the DW factor again rapidly 
saturates to a $T$-independent value. 

\subsection{Temperature evolution of the yield stress}

\begin{figure}
\psfig{file=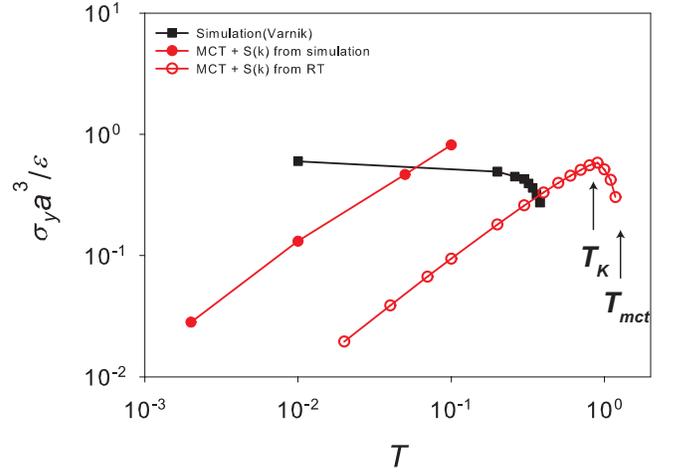,width=8.5cm,clip}
\caption{\label{fig8} 
Temperature dependence of the yield stress obtained from direct
numerical simulation (filled squares) and from MCT
using either the numerically measured structure factor 
(filled circles) or a statistical mechanics approach 
(Replica Theory, open circles). While the simulation results indicate a nearly 
temperature independent yield stress, $\sigma_Y \sim \epsilon / a^3$,
the MCT solution produces instead an `entropic'
yield stress vanishing linearly with $T$ at low $T$. }
\end{figure}

We now analyze the temperature dependence of the yield stress of the 
Lennard-Jones model. 

In Fig.~\ref{fig8}, the yield stress obtained from 
earlier simulations~\cite{yieldLJ,yieldLJ2} and 
from the  MCT equations are plotted as a function of the temperature. 
From the discussion in Sec.~\ref{softflow} for the soft glass, 
we expect the yield stress of the Lennard-Jones system to 
be controlled by the interaction energy between particles, 
and we choose therefore to plot the stress in adimensional units,
$\sigma \to \sigma / (\epsilon / a^3)$, where $\epsilon$ 
represents now the attractive depth of the Lennard-Jones potential.
Using this representation, we find that the numerical results 
for the yield stress are in fact weakly dependent on the temperature, 
rapidly saturating to the $T \to 0$ limit, $\sigma_Y(T=0) / 
(\epsilon / a^3) \sim {\cal O}(1)$, as expected. 
 
Performing the MCT analysis using the low temperature 
structure factor, we find that the predicted yield stress
decreases linearly with the temperature. This is because in this 
regime $S(k)$ is nearly constant, and the MCT equations 
produce an incorrect `entropic' yield stress, i.e. $\sigma_Y \sim k_B T$.  
Finally, using the analytic
structure factor, we again find a yield stress which vanishes
linearly with $T$ at low $T$, with a singular emergence 
near the mode-coupling singularity, mirrorring the
behaviour obtained for the DW factor in Fig.~\ref{fig7}. 

Again, the discrepancy between simulations and MCT predictions 
regarding the physical origin of the yield stress is the same 
as the one uncovered in the above study of the soft glass regime
of harmonic spheres. This shows that this result was not an artefact
of the peculiar harmonic sphere system, nor was it
related to singularities encountered near the jamming transition
in this system. For Lennard-Jones particles, there is 
no jamming singularity in the density regime studied in the 
present section, but similar results are found for this 
well-known glass-forming model system.  

\section{Discussion}

\label{discussion}

We have shown that mode-coupling theory provides 
`first-principles' predictions for the emergence of the 
yield stress in amorphous solids, together with detailed 
predictions for the temperature and density dependences 
of the yield stress in various glassy materials, 
from hard sphere glasses to soft and metallic glasses. 

For hard sphere glasses, the theory correctly predicts the 
emergence of solid behaviour with entropic origin, with a yield 
stress and shear modulus proportional to $k_B T$. The theory 
also predicts a divergence of the yield stress as the random 
close packing density is approached, but the predicted critical 
exponent is too large. We have shown that this is because MCT also
considerably overestimates the degree of localization of the particles
in the glass at rest near the jamming transition.

The theory fares more poorly for both soft glasses and metallic glasses, 
as it again predicts a yield stress proportional to $k_B T$ 
while solidity is in fact the result of direct interparticle 
forces, and scales instead as $\epsilon /a ^3$, where $\epsilon$ is the 
typical energy scale governing particle interactions. 

This means that while the flow curves predicted by MCT for a given material
across the glass transition may have functional forms that are in good 
agreement with the observations, it is not clear whether the 
non-linear flow curves produced in the glass phase are physically
meaningful for particles that cannot be represented 
as effective hard spheres. 

The fact that the mode-coupling theory provides limited insight into 
solid phases is perhaps not surprising, as the theory was initially
developed as an extension of liquid state theories~\cite{gotze}. However, 
since the  theory describes the transformation into the 
arrested glass phase, 
the MCT predictions for the glass dynamics at rest and for the 
glass dynamics under shear flow have been worked out in detail,
and often discussed in connection with experimentally relevant 
questions, such the Boson peak in amorphous systems~\cite{gotze2}, 
and the non-linear flow of glasses~\cite{fuchs4}. 

Our study suggests that one 
should perhaps not try to apply MCT `too deep' into the glass, but 
it must be noted that 
the theory itself can be applied arbitrarily far into the glass 
phase with no internal criterion suggesting that the procedure becomes 
inconsistent, as long as reliable estimates of the static structure factor 
are available.

For Lennard-Jones particles and the soft glass regime 
of harmonic spheres, we have discussed such a criterion. 
We suggested the existence of a temperature
scale $T_q$ below which MCT predictions certainly become unreliable. 
This temperature is such that, below $T_q$,
the averaged static structure 
becomes dominated by the quenched disorder instead of thermal
fluctuations~\cite{criticality}. This implies that MCT predictions
for glassy phases should be more reliable in the regime $T_q < 
T < T_g$. We note, however, 
that MCT only makes crisp predictions near the mode-coupling 
`singularity' $T_{mct}$ (such as square root dependence of the DW factor and 
yield stress) but these are not easy to test since the real system 
is actually in a fluid state at $T_{mct} > T_g$. 

More generally, we believe our work emphasizes the need for more detailed 
theoretical analysis of the non-linear response of amorphous 
solids to external shear flow to produce better theoretical 
understanding of the yield stress in disordered materials. 
Recent progress in the statistical mechanics of the glassy state
using replica calculations~\cite{gr3,mp,PZ,hugo,yoshino} 
provide detailed predictions for 
the thermodynamics, micro-structure, and shear modulus 
of glassy phases, that are, contrary to the MCT result exposed 
in this work, at least consistent with a low-temperature 
harmonic description of amorphous solids. One can hope that an 
these calculations can be extended to treat also the yield 
stress. 

\acknowledgments

We thank G. Biroli and K. Miyazaki for discussions and 
R\'egion Languedoc-Roussillon and 
JSPS Postdoctoral Fellowship for Research Abroad (A. I.) for financial 
support. The research leading to these results has received funding
from the European Research Council under the European Union's Seventh
Framework Programme (FP7/2007-2013) / ERC Grant agreement No 306845.

\end{document}